\documentclass[11pt]{article}
\usepackage[margin=1in]{geometry}
\usepackage{multicol}
\usepackage[T1]{fontenc}
\usepackage[utf8]{inputenc}
\usepackage{authblk}
\usepackage{subcaption}
\usepackage{amsmath,amssymb,amsfonts}
\usepackage{algorithmic}
\usepackage[final]{graphicx}
\usepackage{textcomp}
\usepackage{xcolor}
\usepackage{circuitikz}
\usepackage{adjustbox}
\usepackage{wrapfig}
\usepackage{float} 
\def\BibTeX{{\rm B\kern-.05em{\sc i\kern-.025em b}\kern-.08em
    T\kern-.1667em\lower.7ex\hbox{E}\kern-.125emX}}

\title{Development of an Array of Kinetic Inductance Magnetometers (KIMs)}
\author[1]{Sasha Sypkens}
\author[2]{Farzad Faramarzi}
\author[3]{Marco Colangelo}
\author[1]{Adrian Sinclair}
\author[2]{Ryan Stephenson}
\author[2,4]{Jacob Glasby}
\author[5]{Peter Day}
\author[3]{Karl Berggren}
\author[1,2]{Philip Mauskopf}

\affil[1]{School of Earth and Space Exploration, Arizona State University}
\affil[2]{Department of Physics, Arizona State University}
\affil[3]{Department of Electrical Engineering and Computer Science, Massachusetts Institute of Technology}
\affil[4]{School of Electrical, Computer, and Energy Engineering, Arizona State University}
\affil[5]{Jet Propulsion Laboratory}

\date{}

\begin{document}

\ctikzset{bipoles/thickness=1}

\maketitle

\begin{abstract}
We describe optimization of a cryogenic magnetometer that uses nonlinear kinetic inductance in superconducting nanowires as the sensitive element instead of a superconducting quantum interference device (SQUID). The circuit design consists of a loop geometry with two nanowires in parallel, serving as the inductive section of a lumped LC resonator similar to a kinetic inductance detector (KID).  This device takes advantage of the multiplexing capability of the KID, allowing for a natural frequency multiplexed readout.  The Kinetic Inductance Magnetometer (KIM) is biased with a DC magnetic flux through the inductive loop.  A perturbing signal will cause a flux change through the loop, and thus a change in the induced current, which alters the kinetic inductance of the nanowires, causing the resonant frequency of the KIM to shift.  This technology has applications in astrophysics, material science, and the medical field for readout of Metallic Magnetic Calorimeters (MMCs), axion detection, and magnetoencephalography (MEG). 
\end{abstract}

\newpage

\vspace{11pt}
\noindent \Large\textbf{Introduction}

Highly sensitive magnetic sensors are useful in many fields, particularly in astronomy, medicine, and geology. \cite{squidgeophys,squidbiomag,clarke2006squid} Currently, the Superconducting QUantum Interference Device (SQUID) is the most sensitive magnetometer, with a sensitivity of $fT / \sqrt{Hz}$ \cite{clarke2006squid}. Arrays of SQUIDs have been used for time division multiplexing (TDM) of superconducting detectors for over a decade \cite{irwin2002}.  However, scaling to the increasing number of detectors used in modern instruments is challenging partly because the SQUID requires multiple layers during fabrication, which increases time and cost to make arrays \cite{kpup}.  A recent development called $\mu$MUX attempts to couple each SQUID to a corresponding transmission line resonator to achieve frequency domain multiplexing (FDM) with a high multiplexing factor.  This uses an RF SQUID coupled to a modulation coil to counteract the periodic nature of the device, and a load inductor to couple to the resonator for readout. Each resonator has a unique resonant frequency, and is shifted by the change in inductance in the RF SQUID \cite{uMUX} \cite{doi:10.1063/1.1791733} \cite{doi:10.1063/1.4829156}.

In addition to improving SQUID readout, there has been some publication on using Dayem bridges instead of Josephson Junctions. \cite{doi:10.1063/1.4948477} Using a Dayem bridge and a small SQUID diameter, one group was able to achieve 50$n\Phi_0/\sqrt{Hz}$ \cite{SQUIDtinyFLUX}  Furthermore, it has been shown that kinetic inductance rather than magnetic inductance can modulate nanoSQUIDs, effectively reducing the size of the device by an order of magnitude.  \cite{nanoSQUID}

Alternatively, two groups independently developed a superconducting magnetic sensor that uses nanowires with high kinetic inductance instead of Josephson Junctions as the sensing component \cite{kpup} \cite{KIM_Nature}.  These devices are similar to the Lumped Element Kinetic Inductance Detector (LEKID) in that they are planar lumped element devices and use superconductors which have high nonlinear kinetic inductance.  However, they have not been able to achieve SQUID sensitivity or develop an array of magnetometers.  This paper aims to demonstrate an array of KIMs read out using FDM and a discussion on its sensitivity.  The KIM can be used for magnetic field sensing, direct readout of magnetic x-ray detectors, and current-coupled readout. \cite{kpup} \cite{KIM_Nature} \cite{Asfaw_2018}

\vspace{11pt}
\noindent \Large\textbf{Operating Principles}

The kinetic inductance of a nonlinear superconducting thin film in which the thickness, $t$ is much less than the penetration depth, $\lambda$ can be expressed by

\begin{equation}
    L_k = L_k(0) \left(1+\frac{I_b^2}{I_*^2}+... \right),
\end{equation}

\noindent where $I_*$ is the characteristic current dependent on the material and geometry of the film that sets the nonlinearity of the film, and $L_k(0)$ is the kinetic inductance of the superconductor with no bias current, $I_b$.  The superconductor is at its peak nonlinearity when biased close to $I_*$.

When formed into a loop, the superconductor obeys the following expression:

\begin{equation}
    \Phi_b - LI_s = n\Phi_0,
\end{equation}

\noindent in which $I_s$ is the shielding current induced by the bias field, the total inductance, $L = L_k + L_g$ includes the geometric term and the kinetic inductance of the material, $\Phi_b$ is the bias flux, and $n$ is an integer multiple of $\Phi_0$, the flux quantum.  When the loop is biased with a magnetic flux perpendicular to its plane, the superconductor generates an opposing shielding current, which from looking at Eq. 1, determines the kinetic inductance.  If there is a perturbation in the flux, the kinetic inductance of the loop will change.  When the loop is incorporated into a resonator, the change in kinetic inductance can be seen as a shift in resonance, $\omega_r = 1/\sqrt{LC}$. \cite{kpup} \cite{KIM_Nature} \cite{Asfaw_2018}

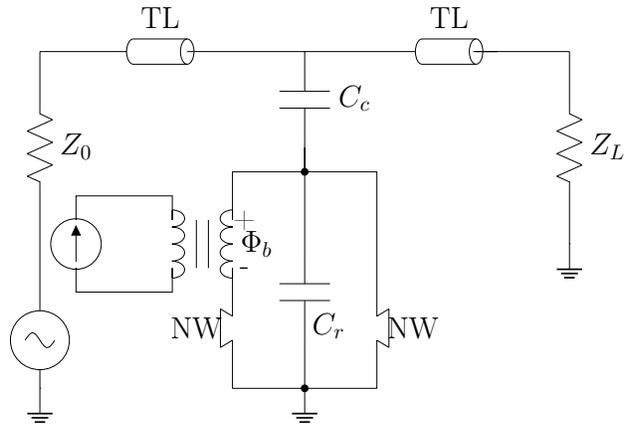
\begin{wrapfigure}{r}{0.5\linewidth}
\centering
\begin{adjustbox}{scale=0.8}
\begin{circuitikz}[scale=0.8,american] 
\draw

(-2,2) -- (-0.5,2)
(-0.5,2) to [TL=TL] (1.5,2)
(1.5,2) -- (5.5,2)
(5.5,2) to [TL=TL] (7.5,2)
(7,2) -- (9,2)
(3.5,2) to [capacitor = $C_c$] (3.5,0)
(3.5,0) -- (3.5,-2)
(3.5,0) to[short, -*] (3.5,-0.5)
(2,-0.5) -- (5,-0.5)
(3.5,-2) to [capacitor] (3.5,-4)
(3.5,-4)to[short, -*] (3.5,-5)
(4,-3.7) node{$C_r$}

(5.75,-3.75) node{NW}
(5,-0.5) -- (5,-3.5)
(5.25,-3.35) -- (5.25,-4.15)
(5.25,-3.35) -- (5,-3.5)
(5.25,-4.15) -- (5,-4)
(5,-4) -- (5,-5)

(1.25,-3.75) node{NW}
(2,-0.5) -- (2,-1.5)
(2,-2.5) -- (2,-3.5)
(1.75,-3.35) -- (1.75,-4.15)
(1.75,-3.35) -- (2,-3.5)
(1.75,-4.15) -- (2,-4)
(2,-4) -- (2,-5)

(2,-2.5) to [american inductor] (2,-1.5)
(2.25,-2.5) node{-}
(2.5,-2) node{$\Phi_b$}
(2.25,-1.5) node{+}
(0.75,-1.5) to [american inductor] (0.75,-2.5)
 (-1.25, -2.5) to [isource] (-1.25,-1.5)
 (-1.25,-1.5) -- (-1.25,-1) -- (0.75,-1) -- (0.75,-1.5)
 (-1.25,-2.5) -- (-1.25,-3) -- (0.75,-3) -- (0.75,-2.5)
 (1.25,-1.5) -- (1.25,-2.5)
 (1.5,-1.5) -- (1.5,-2.5)

(2,-5) -- (5,-5)
(3.5,-5) to node[ground]{} (3.5,-5)
(-2,-3) to [vco] (-2,-5)
(-2,2) -- (-2,1)
(-2,1) to [resistor=$Z_0$] (-2,-1)
(-2,-1) -- (-2,-3)
(-2,-5) to node[ground]{} (-2,-5)
(9,2) -- (9,1)
(9,1) to [resistor=$Z_L$] (9,-1)
(9,-1) -- (9,-2)
(9,-2) to node[ground]{} (9,-2)
;

\end{circuitikz}
\end{adjustbox}
\caption{Circuit diagram of a KIM}
\label{1}
\vspace{50pt}
\end{wrapfigure}

\vspace{11pt}
\noindent \Large\textbf{Experiment}

\vspace{11pt}
\noindent \large\textit{Design}

The KIM is a microwave resonator coupled to a CPW transmission line via an interdigitated coupling capacitor.  The resonator consists of two nanowires acting as inductors in parallel with an interdigitated resonant capacitor.  Both capacitors are interdigitated to reduce two-level system (TLS) noise and the geometry of the nanowires are kept small ( $w \approx$ 50nm, $t \approx$ 20nm) to maximize $\alpha$, the kinetic inductance fraction of the device.  TLS noise originates from coupling between the resonator and electric dipole moments on the substrate/superconductor interface. \cite{Noroozian_2009} \cite{doi:10.1063/1.2711770} There are four chips, each with an array of 8 resonators, where the resonant capacitor varies in each device.  NbN was chosen to be the device material due to its high kinetic inductance.  The nanowire length for each array was chosen to be 250nm, 500nm, 750nm, and 1$\mu$m.


\begin{wrapfigure}{l}{0.5\linewidth}
\centering
\subfloat[optical micrograph]{\includegraphics[width = 1.8in]{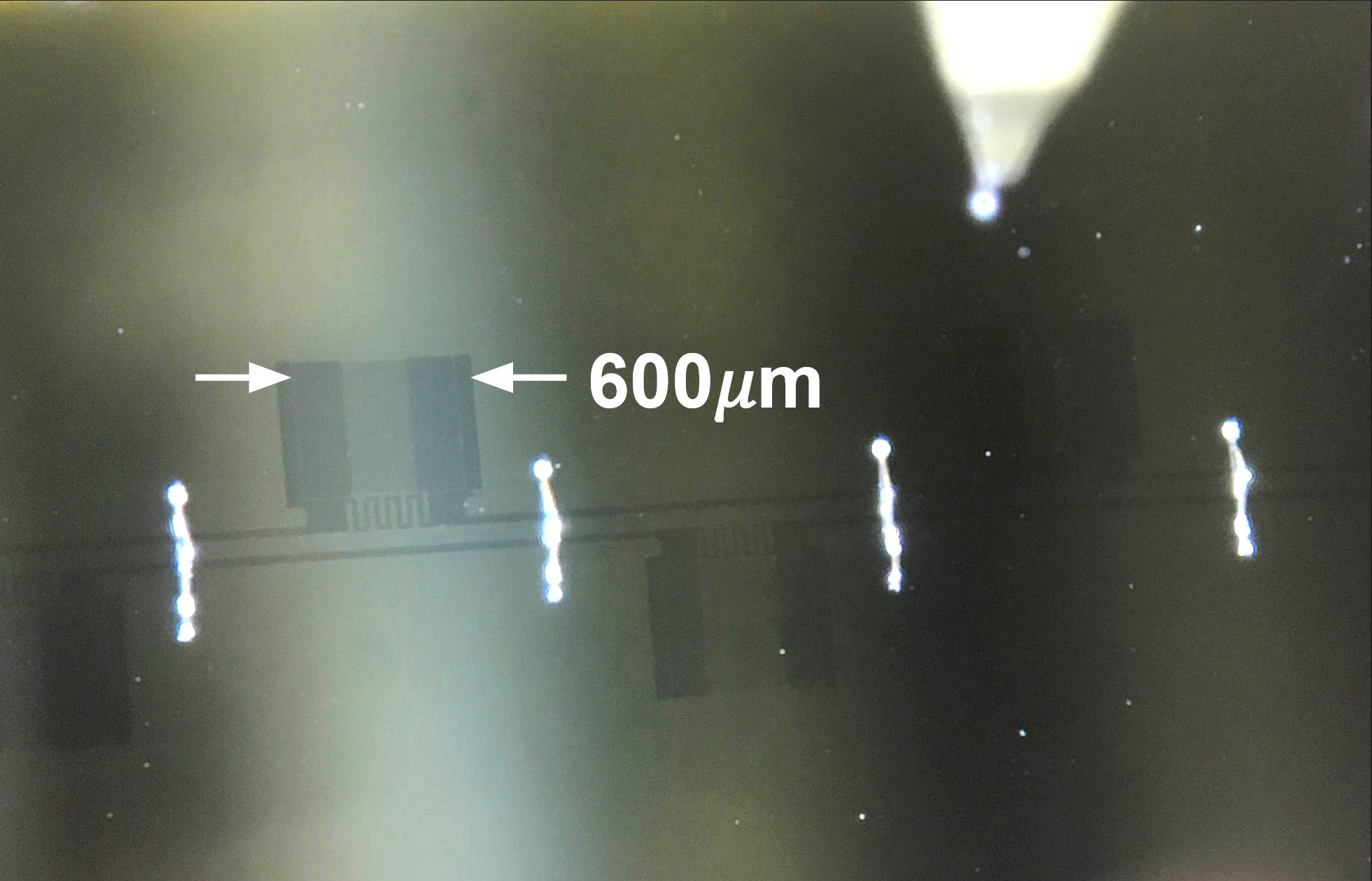}} \\
\vspace{5pt}
\subfloat[SEM of one resonator]{\includegraphics[width = 3.1in]{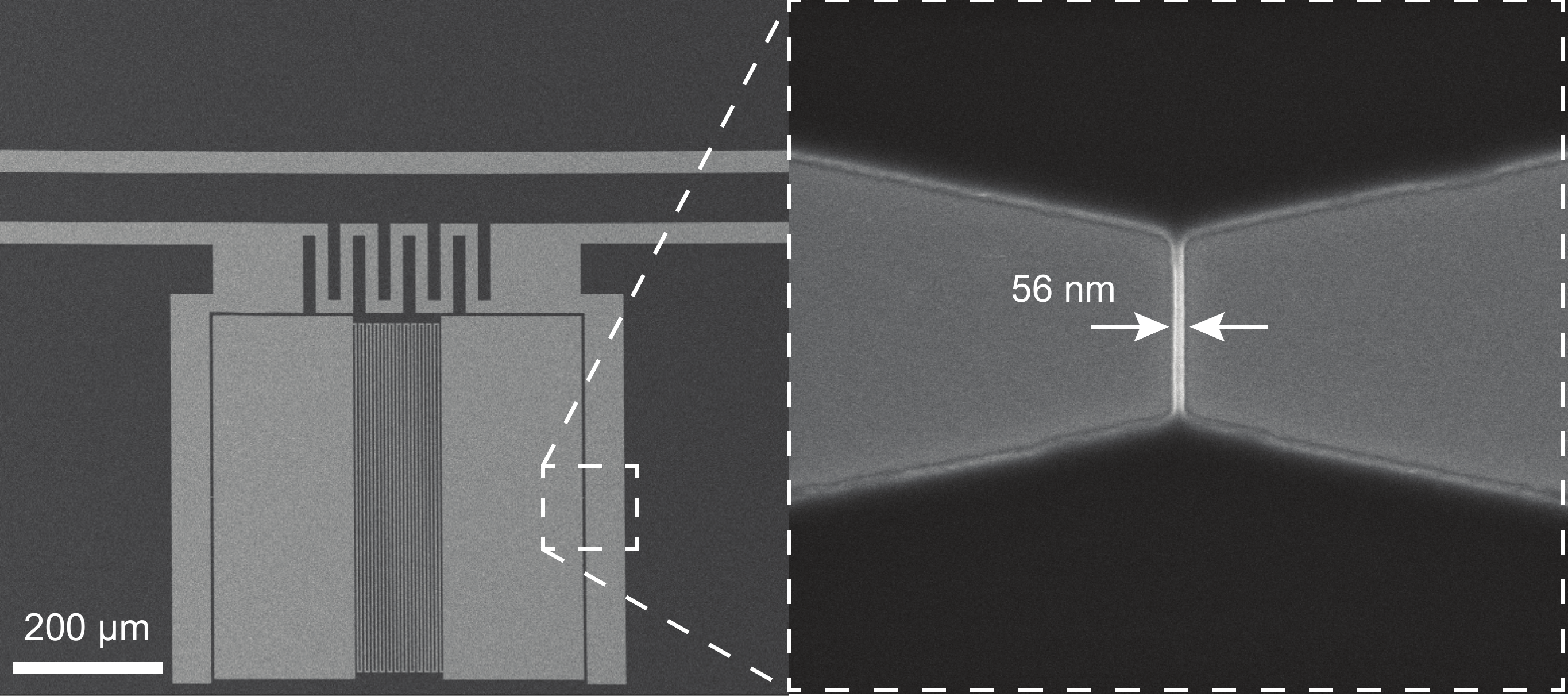}}
\caption{Optical and scanning electron micrographs of the fabricated device.  The dark portion in b) is NbN and the light portion is the substrate.  The nanowire is also highlighted in white in the second SEM image.}
\label{device}
\vspace{-10pt}
\end{wrapfigure}

\vspace{11pt}
\noindent \large\textit{Fabrication} 

The KIM arrays were fabricated from a $\approx 16.8\,\mathrm{nm}$-thick niobium nitride (NbN) film. NbN was reactively sputtered at room temperature on a high-resistivity silicon wafer \cite{dane2017bias}. The room-temperature sheet resistance was $R_{\mathrm{s}}\approx 165\,\mathrm{\Omega}$ per square, the critical temperature was $T_{\mathrm{c}}=8.74\,\mathrm{K}$, and the critical current was roughly 50$\mu$A. The arrays were patterned using $125\,\mathrm{kV}$-electron beam lithography with ZEP530A positive tone resist. The patterns were transferred into NbN with CF$_4$ reactive ion etching (more fabrication details are reported in previous publications
on superconducting nanowire single photon detectors \cite{najafi2014fabrication,yang2005fabrication}). Fig.~\ref{device} shows optical and scanning electron micrographs of the fabricated device.


\vspace{11pt}
\noindent \large\textit{Test Setup}

An in-house Helmholtz coil made with superconducting wire and calibrated with a commercial hall sensor was used to bias the 

\begin{wrapfigure}{l}{0.5\linewidth}
\centering
\begin{adjustbox}{scale=0.68}
\begin{circuitikz}[scale=0.68]

    \draw
    (0,0) -- (3,0)
    (1,-1) to [R] (1,1)
    (3,0) to [TL = Coax] (4,0)
    (4,0) -- (6,0)
    (6,0) to [twoport, t=DUT] (7,0)
    (7,0) -- (9,0)
    (9,0) to [TL = Coax] (10,0)
    (10,0) -- (12,0)
    (12,0) to [amp] (13,0)
    (13,0) -- (14,0)
    
    (0,0) -- (0,-2)
    (0,-2) to [TL = Coax] (0,-3)
    (0,-3) -- (0,-4)
    (0,-4) to [C=DC Block] (0,-5)
    (0,-5) -- (0,-6)
    (0,-6) to [TL = Coax] (0,-7)
    (0,-7) -- (0,-8)
    
    (14,0) -- (14,-2)
    (14,-3) to [TL = Coax] (14,-2)
    (14,-3) -- (14,-4)
    (14,-5) to [C=DC Block] (14,-4)
    (14,-5) -- (14,-6)
    (14,-7) to [TL = Coax] (14,-6)
    (14,-7) -- (14,-8)    
    
    (0,-8) -- (6,-8)
    (6,-8) to [twoport, t=VNA] (7,-8)
    (7,-8) -- (9,-8)
    (10,-8) to [amp] (9,-8)
    (10,-8) -- (12,-8)
    (12,-9) to [R] (12,-7)
    (12,-8) -- (14,-8);
    
    \draw[orange,thick,dashed] (-2,3) rectangle (16,-9)
    (6.5,-9.5) node[]{300K};

    \draw[purple,thick,dashed] (-1.5,2.5) rectangle (15.5,-5.5)
    (6.5,-6) node[]{40K};    
    
    \draw[blue,thick,dashed] (-1,2) rectangle (15,-3.5)
    (6.5,-4) node[]{4K}; 
    
    \draw[black,thick,dashed] (5,1.5) rectangle (8,-1)
    (6.5,-1.5) node[]{Magnetic Shield}; 
    
    \draw
    (7.5,1) node[]{$\Phi_b$};

\end{circuitikz}
\end{adjustbox}
\caption{Test Setup of the DUT with a magnetic flux bias, $\Phi_b$.  The coaxial cables use a combination of stainless steel (300K to 3K stages) and copper (outside of the cryostat and at the 3K stage) cables.  The cold amplifier is a lab made LNA with 10K noise temperature and 30dB of gain, and the room temperature amplifier is from mini-circuits.}
\label{2}

\end{wrapfigure}
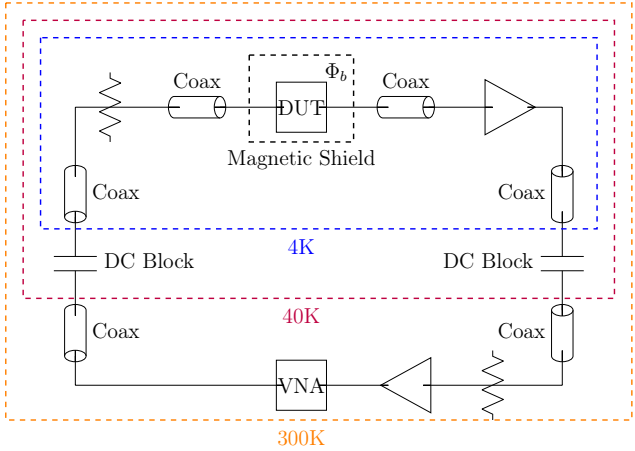

\noindent resonators.  The Helmholtz coil and device were wrapped with Nb foil on the 3K testbed to shield them from environmental magnetic fields.  The magnetic bias and shield are indicated next to the DUT in Fig.~\ref{2}.  To read out the KIM array, an attenuator and low noise amplifier (LNA) are placed on either side of the DUT, with coax cables leading on either end to the VNA.  The DC blocks shown in the same figure are inner/outer blocks to act as thermal breaks.

\vspace{11pt}
\noindent \Large\textbf{Results and Discussion}

\vspace{11pt}
\noindent \large\textit{Temperature Response}

\begin{wrapfigure}{l}{0.5\linewidth}
\centering
\includegraphics[width = 3in]{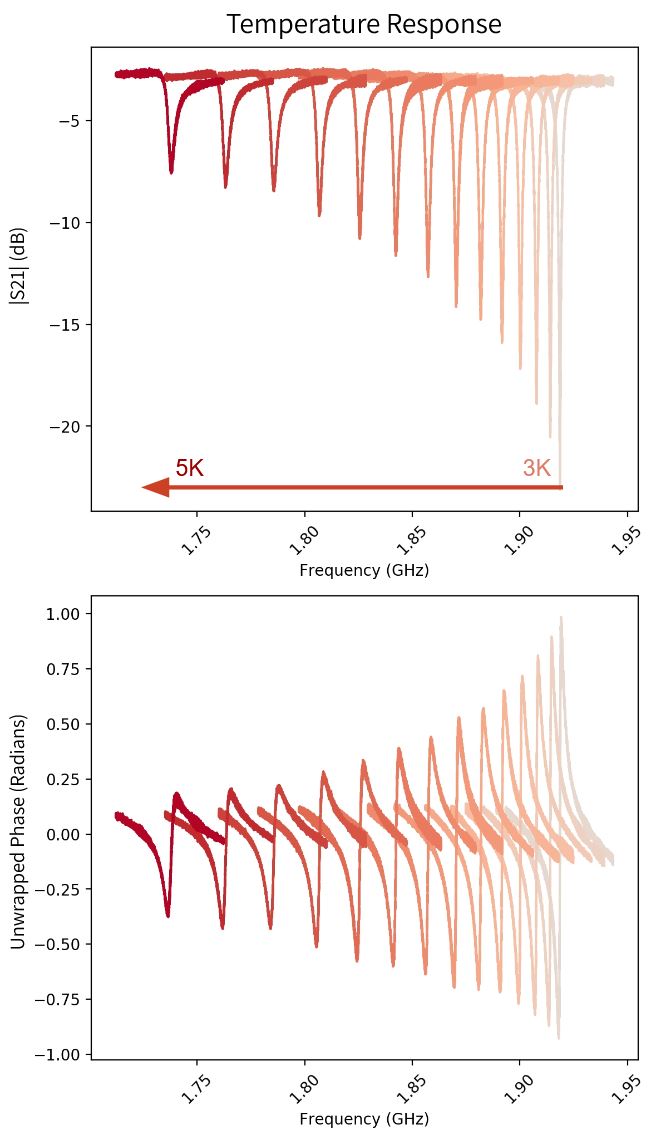}
\caption{Response of one of the resonators in the 1-$\mu m$-long array as a function of temperature going from 3K to 5K.  The top image shows $S_{21}$ and the lower image shows the response in phase space.}
\label{temp}
\vspace{20pt}
\end{wrapfigure}

Fig.~\ref{temp} shows the temperature response of a resonator from the 1$\mu$m array with a -85dBm input power at the device. \cite{Carter2016} This behavior is expected because it is seen from measurements taken from a typical LEKID architecture.  At the lowest temperature (3K), the internal quality factor for all of the resonator arrays was around 4,000 with the total quality factor heavily dominated  

\noindent by the coupling.

\vspace{11pt}
\noindent \large\textit{Magnetic Response}

The responsivity of a KIM is how much the resonant frequency shifts with a magnetic field bias, $df/fdB$.  As stated previously, the resonant frequency shifts as a function of inductance, and so 

\begin{equation}
    \bigg( \frac{1}{f}\frac{df}{dB} \bigg) \propto -\frac{1}{2}\frac{\delta L}{L} = - \frac{\alpha}{2}\frac{\delta L_k}{L_k},
\end{equation}

\noindent where $\alpha$ is the kinetic inductance fraction, $L_k/(L_k+L_g)$.  The kinetic inductance term can be seen as the kinetic energy divided by the pairing energy \cite{Zmuidzinas-2012},

\begin{equation}
    \frac{E_k}{E_p} = \frac{\frac{1}{2}L_kI^2}{2N_0 \Delta_0^2 V},
\end{equation}

\noindent in which $N_0$ is the single-spin density of states, $\Delta_0$ is the band gap energy of the superconductor at T=0, and $V$ is the volume of the nanowires.  Solving for $I_s$ in equation (2), taking into account there being two nanowires, and plugging it in to the equation above, the responsivity becomes

\begin{equation}
    R = \frac{\alpha L_k}{4N_0 \Delta_0^2 V} \bigg( \frac{\Phi_b - n\Phi_0}{2L_k} \bigg)^2 .
\end{equation}

\noindent Doing some simple term cancellation shows that $L_k$ should be high enough for $\alpha \approx 1$ but not more than necessary because it is inversely proportional to responsivity.  

The above equation also shows that once $\Phi_b = n\Phi_0$, responsivity goes to 0, indicating that the nanowires have turned normal and at that point in time, 174 flux quanta enter the loop before the nanowires turn superconducting again.  This behavior is periodic in that the response starts over after each nanowire-turn-normal event.

\begin{figure*}[h]
\centering
\subfloat[Magnetic Response Fit: 1$\mu$m long nanowire]{\includegraphics[height = 2in]{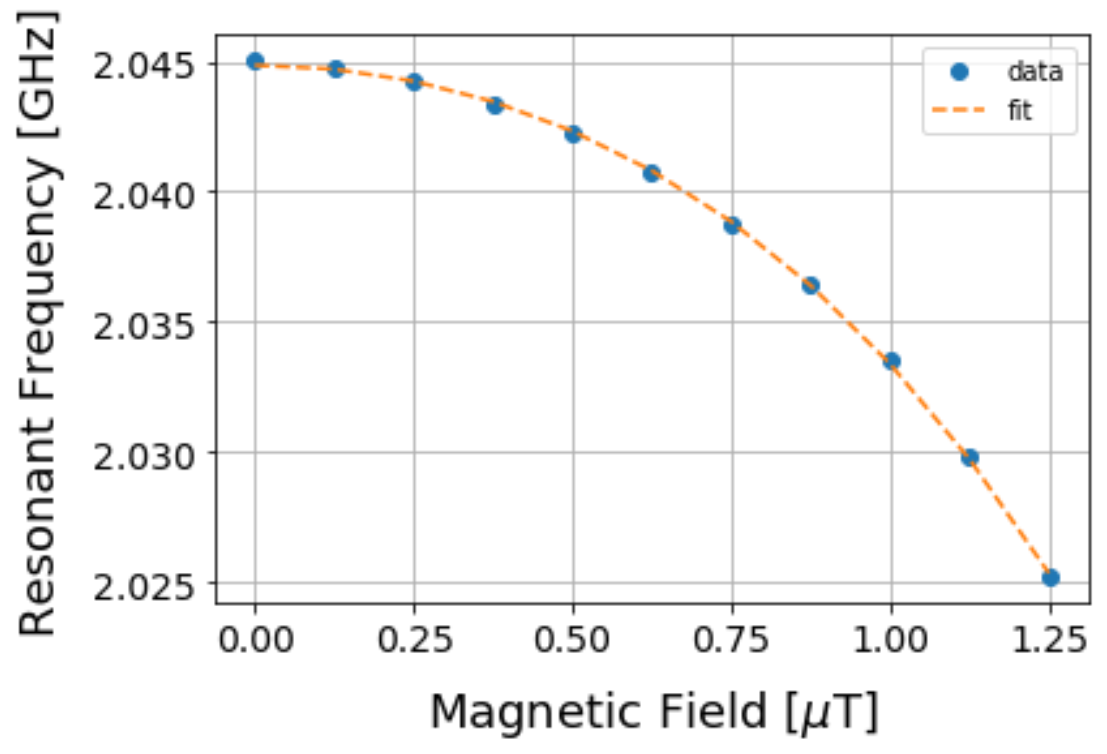}}
\hspace{15pt}
\subfloat[Calculated Responsivity and Flux Noise]{\includegraphics[height = 2in]{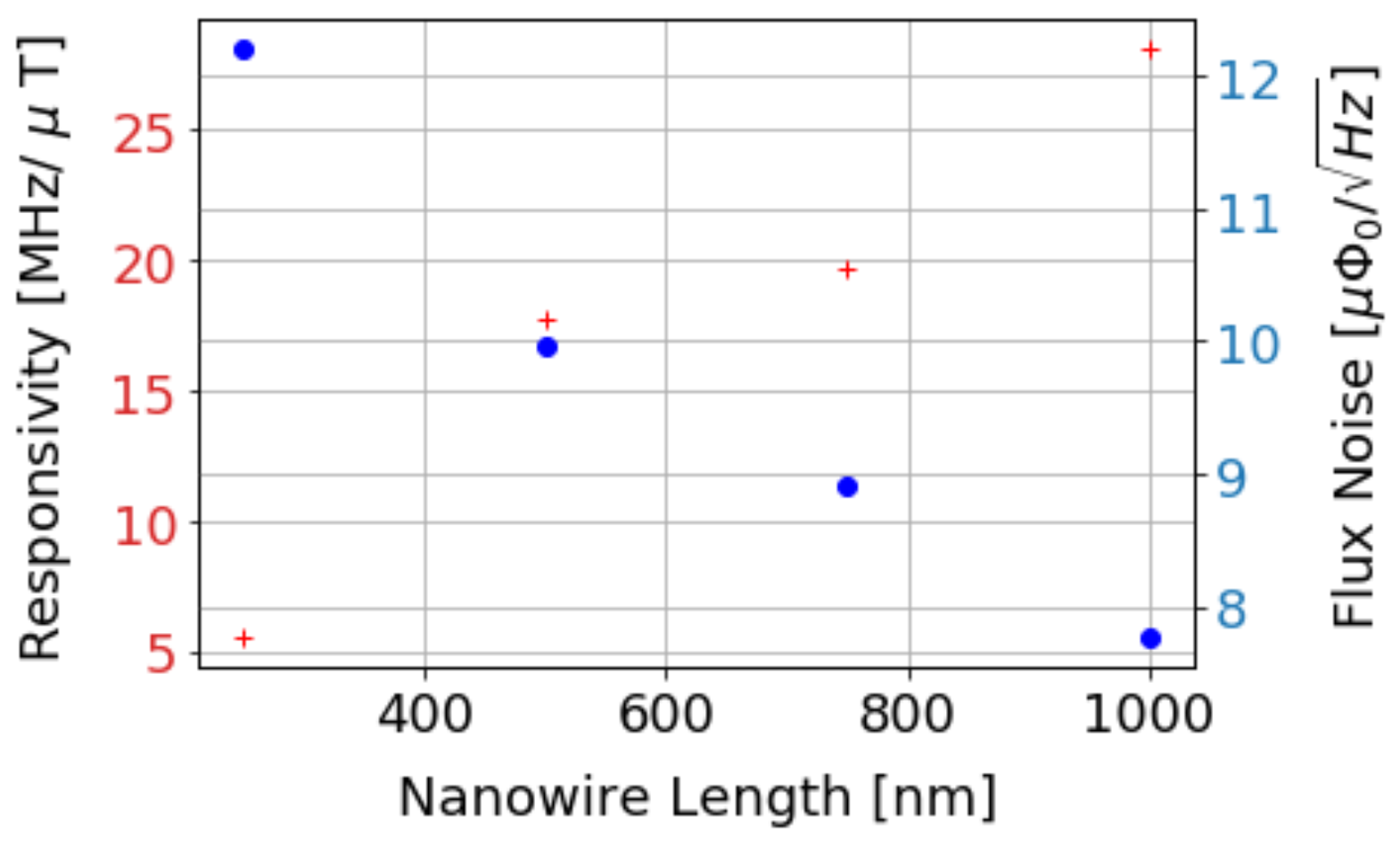}} 
\caption{KIM data. a) magnetic response with fit of resonator 3 in the 1-$\mu m$-long array}, b) calculated responsivity when biased near $I_*$ (red + sign) and flux noise (blue dot) for all arrays (250nm, 500nm, 750nm, 1$\mu$m long nanowires).
\label{magres}
\end{figure*}

The magnetic response of one resonator and a comparison of responsivity and calculated flux noise based on nanowire length are plotted in Fig.~\ref{magres}.  In image (a), the resonant frequency with no magnetic field bias is 2.045GHz, and the resonant frequency shifts downward as the magnetic field is increased.  At 1.25$\mu$T, the resonant frequency shifts maximally, and if biased further would shift back up to the original value, indicating the nanowires turn normal and let in some flux quanta before turning superconducting, again.  Based on the geometry of the loop, 174 flux quanta enter the loop at each period.  

The magnetic response fit equation in image (a) is

\begin{equation}
    f_r = - \frac{B^2}{a^2} - \frac{B^4}{b^4},
\end{equation}

\noindent where $a=9.7nT/GHz$ and $b=1.8nT/GHz$ are the fit parameters and $f_r$ is the resultant resonant frequency from the magnetic field bias. \cite{kpup}  The nonlinearity from Eq. 1 can be seen in the responsivity at higher applied magnetic field.

In Fig.~\ref{magres}, image (b) compares the response of all four arrays, one being for calculated flux noise and the other for responsivity.  It can be seen that flux noise is inversely proportional and responsivity is directly proportional to nanowire length.  This is because the nanowire's zero-bias kinetic inductance increases with length, thereby decreasing responsivity.  This in turn corresponds to lower noise because it is inversely proportional to responsivity, as can be seen by calculating the equivalent total flux noise power spectral density at the input of the KIM, 

\begin{equation}
    S_B = \bigg( \frac{1}{f} \frac{df}{dB} \bigg)^{-2} \frac{k_BT_a}{8Q^2P},
\end{equation}

\noindent where $k_B$ is the Boltzmann constant, $T_a$ is the amplifier noise temperature, $Q$ is the quality factor of the resonator, $P$ is the input power, and $\frac{df}{dB}$ is the responsivity calculated from the magnetic response fit in image (a). One should note that the calculation assumes the system is dominated by the amplifier noise because other sources such as TLS noise is considerably smaller from the use of IDCs.  \cite{kpup} The calculations in Fig.~\ref{magres} (b) for the flux noise used an input power of -79dB.  From there, the flux noise can be easily converted from magnetic field noise.  For the 1$\mu$m array, the flux noise was calculated to be roughly 7.5$\mu\Phi_0/\sqrt{Hz}$.

\vspace{11pt}
\noindent \Large\textbf{Conclusion and Future Work}

Noise measurements were taken using the Keysight E5072A VNA by setting the center frequency to a 0Hz span and the IFBW to 100kHz for a nyquist bandwidth of 50 kHz.  Unfortunately, the data shows that the system was dominated by the noise of the VNA.  To mitigate this effect from future noise measurements, more amplification will be used after the cold amplifier than the 13dB warm amplifier that was used for this measurement.  Additionally, we will use a ROACH2 with MUSIC-DAC/ADC readout system \cite{blast} and an amplifier with lower noise temperature than the one that was used for this measurement (10K noise temperature).

Of the resonators measured, the highest responsivity was in the 1$\mu$m array at 28MHz/$\mu$T, corresponding to a calculated flux noise of 7.5$\mu\Phi_0/\sqrt{Hz}$.  To compare to a SQUID, its noise would need to be 1$\mu\Phi_0/\sqrt{Hz}$. \cite{kpup}  

In this experiment, the thickness of the entire resonator, both the nanowires and the remaining components, was the same at roughly 17nm.  To ensure the device is only affected by the nanowires, a different fabrication process needs to be implemented so that the nanowires are thin and everything else is either thicker or a different material that does not have high kinetic inductance.  Optimizing the design in this way will reduce the flux noise to lower than 7$\mu\Phi_0/\sqrt{Hz}$.  Furthermore, the average internal quality factor of these resonators was around 4,000, with the coupling factor reducing the total quality factor by an order of magnitude.  Although the coupling needs to change for a higher total quality factor, measuring these resonators at 800mK instead of 3K would greatly improve their quality.  Lastly, the magnetic shield was just Nb foil wrapped around the device and Helmholtz coil, which led to large gaps in shielding at the input and output of the DUT.  Another goal is to design a proper magnetic shield for the setup.


\bibliographystyle{unsrt} 
\bibliography{references.bib}

\begin{thebibliography}{10}

\bibitem{squidgeophys}
Harold Weinstock and Jr. Overton, William~C.
\newblock {\em {SQUID Applications to Geophysics}}.
\newblock Society of Exploration Geophysicists, 01 1981.

\bibitem{squidbiomag}
Rainer K{\"o}rber, {Jan Hendrik} Storm, Hugh Seton, {Jyrki P.} Makela, Ritva
  Paetau, Lauri Parkkonen, Christoph Pfeiffer, Bushra Riaz, {Justin F.}
  Schneiderman, Hui Dong, {Seong Min} Hwang, Lixing You, Ben Inglis, John
  Clarke, {Michelle A.} Espy, {Risto J.} Ilmoniemi, {Per E.} Magnelind, {Andrei
  N.} Matlashov, {Jaakko O.} Nieminen, {Petr L.} Volegov, {Koos C J}
  Zevenhoven, Nora H{\"o}fner, Martin Burghoff, Keiji Enpuku, {S. Y.} Yang,
  {Jen Jei} Chieh, Jukka Knuutila, Petteri Laine, and Jukka Nenonen.
\newblock Squids in biomagnetism: A roadmap towards improved healthcare.
\newblock {\em Superconductor Science and Technology}, 29(11), September 2016.

\bibitem{clarke2006squid}
John Clarke and Alex~I. Braginski.
\newblock {\em The SQUID Handbook Fundamentals and Technology of SQUIDs and
  SQUID Systems}, volume~1.
\newblock Wiley-VCH, Weinheim, 2006.

\bibitem{irwin2002}
K.~D. Irwin, L.~R. Vale, N.~E. Bergren, S.~Deiker, E.~N. Grossman, G.~C.
  Hilton, S.~W. Nam, C.~D. Reintsema, D.~A. Rudman, and M.~E. Huber.
\newblock Time-division squid multiplexers.
\newblock {\em AIP Conference Proceedings}, 605(1):301--304, 2002.

\bibitem{kpup}
Aditya~Shreyas Kher.
\newblock {\em Superconducting nonlinear kinetic inductance devices}.
\newblock PhD thesis, California Institute of Technology, 2017.

\bibitem{uMUX}
Sebastian Kempf, Mathias Wegner, Andreas Fleischmann, Loredana Gastaldo, Felix
  Herrmann, Maximilian Papst, Daniel Richter, and Christian Enss.
\newblock Demonstration of a scalable frequency-domain readout of metallic
  magnetic calorimeters by means of a microwave squid multiplexer.
\newblock {\em AIP Advances}, 7(1):015007, 2017.

\bibitem{doi:10.1063/1.1791733}
K.~D. Irwin and K.~W. Lehnert.
\newblock Microwave squid multiplexer.
\newblock {\em Applied Physics Letters}, 85(11):2107--2109, 2004.

\bibitem{doi:10.1063/1.4829156}
Omid Noroozian, John A.~B. Mates, Douglas~A. Bennett, Justus~A. Brevik,
  Joseph~W. Fowler, Jiansong Gao, Gene~C. Hilton, Robert~D. Horansky, Kent~D.
  Irwin, Zhao Kang, Daniel~R. Schmidt, Leila~R. Vale, and Joel~N. Ullom.
\newblock High-resolution gamma-ray spectroscopy with a microwave-multiplexed
  transition-edge sensor array.
\newblock {\em Applied Physics Letters}, 103(20):202602, 2013.

\bibitem{doi:10.1063/1.4948477}
M.~Arzeo, R.~Arpaia, R.~Baghdadi, F.~Lombardi, and T.~Bauch.
\newblock Toward ultra high magnetic field sensitivity yba2cu3o7 nanowire based
  superconducting quantum interference devices.
\newblock {\em Journal of Applied Physics}, 119(17):174501, 2016.

\bibitem{SQUIDtinyFLUX}
Denis Vasyukov, Yonathan Anahory, Lior Embon, Dorri Halbertal, Jo~Cuppens, Lior
  Neeman, Amit Finkler, Yehonathan Segev, Yuri Myasoedov, Michael Rappaport,
  Martin Huber, and Eli Zeldov.
\newblock A scanning superconducting quantum interference device with single
  electron spin sensitivity.
\newblock {\em Nature nanotechnology}, 8, 09 2013.

\bibitem{nanoSQUID}
Adam McCaughan, Qingyuan Zhao, and Karl Berggren.
\newblock Nanosquid operation using kinetic rather than magnetic induction.
\newblock {\em Scientific Reports}, 6:28095, 06 2016.

\bibitem{KIM_Nature}
Juho Luomahaara, Visa Vesterinen, Leif Gr{\"o}nberg, and Juha Hassel.
\newblock Kinetic inductance magnetometer.
\newblock {\em Nature Communications}, 5, 2014.
\newblock Project code: 72785.

\bibitem{Asfaw_2018}
A.~T. Asfaw, E.~I. Kleinbaum, T.~M. Hazard, A.~Gyenis, A.~A. Houck, and S.~A.
  Lyon.
\newblock Skiffs: Superconducting kinetic inductance field-frequency sensors
  for sensitive magnetometry in moderate background magnetic fields.
\newblock {\em Applied Physics Letters}, 113(17):172601, Oct 2018.

\bibitem{Noroozian_2009}
Omid Noroozian, Jiansong Gao, Jonas Zmuidzinas, Henry~G. LeDuc, Benjamin~A.
  Mazin, Betty Young, Blas Cabrera, and Aaron Miller.
\newblock Two-level system noise reduction for microwave kinetic inductance
  detectors.
\newblock 2009.

\bibitem{doi:10.1063/1.2711770}
Jiansong Gao, Jonas Zmuidzinas, Benjamin~A. Mazin, Henry~G. LeDuc, and Peter~K.
  Day.
\newblock Noise properties of superconducting coplanar waveguide microwave
  resonators.
\newblock {\em Applied Physics Letters}, 90(10):102507, 2007.

\bibitem{dane2017bias}
Andrew~E Dane, Adam~N McCaughan, Di~Zhu, Qingyuan Zhao, Chung-Soo Kim, Niccolo
  Calandri, Akshay Agarwal, Francesco Bellei, and Karl~K Berggren.
\newblock Bias sputtered nbn and superconducting nanowire devices.
\newblock {\em Applied Physics Letters}, 111(12):122601, 2017.

\bibitem{najafi2014fabrication}
Faraz Najafi, Andrew Dane, Francesco Bellei, Qingyuan Zhao, Kristen~A Sunter,
  Adam~N McCaughan, and Karl~K Berggren.
\newblock Fabrication process yielding saturated nanowire single-photon
  detectors with 24-ps jitter.
\newblock {\em IEEE Journal of Selected Topics in Quantum Electronics},
  21(2):1--7, 2014.

\bibitem{yang2005fabrication}
Joel~KW Yang, Eric Dauler, Antonin Ferri, Aaron Pearlman, Aleksandr Verevkin,
  Gregory Gol'tsman, Boris Voronov, Roman Sobolewski, William~E Keicher, and
  Karl~K Berggren.
\newblock Fabrication development for nanowire ghz-counting-rate single-photon
  detectors.
\newblock {\em IEEE Transactions on Applied Superconductivity}, 15(2):626--630,
  2005.

\bibitem{Carter2016}
F.~W. Carter, T.~S. Khaire, V.~Novosad, and C.~L. Chang.
\newblock scraps: An open-source python-based analysis package for analyzing
  and plotting superconducting resonator data.
\newblock {\em IEEE Transactions on Applied Superconductivity}, 27(4):1--5,
  June 2017.

\bibitem{Zmuidzinas-2012}
Jonas Zmuidzinas.
\newblock Superconducting microresonators: Physics and applications.
\newblock {\em Annual Review of Condensed Matter Physics}, 3, 2012.

\bibitem{blast}
Samuel Gordon et~al.
\newblock An open source, fpga-based lekid readout for blast-tng: Pre-flight
  results.
\newblock {\em Journal of Astronomical Instrumentation}, 5(04):1641003, 2016.

\end{thebibliography}

\end{document}